\documentclass{aa}  

\usepackage{color}

\usepackage{graphicx}
\usepackage{txfonts}

\usepackage{amssymb}
\usepackage{multirow}
\usepackage{pifont}
\usepackage{diagbox}

\makeatletter
    
    \newcommand{\Rmnum}[1]{\expandafter\@slowromancap\romannumeral #1@}
\makeatother
\begin{document}

      
      \title{Constructing a refined model of small bodies in the solar system -- \Rmnum2. The Plutinos}


        \author{Yue Chen
          \inst{1,2}
          \and
          Jian Li\inst{1,2} 
          }

   \institute{School of Astronomy and Space Science, Nanjing University, 163 Xianlin Avenue, Nanjing 210023, PR China\\
              \email{ljian@nju.edu.cn}
         \and
         Key Laboratory of Modern Astronomy and Astrophysics in Ministry of Education, Nanjing University, Nanjing 210023, PR China
             }

 
  \abstract
   {As the second part of our study, in this paper, we proceed to refine the solar system model by incorporating the gravitational influence of Plutinos in Neptune's 2:3 mean motion resonance (MMR).}
   {We aim to develop the arc model to represent the global perturbation of Plutinos by taking into account their asymmetric spatial distribution resulting from the 2:3 MMR, and demonstrate the difference to the commonly employed ring model.}
   {The global perturbation of Plutinos is measured by the change in the Sun-Neptune distance. We begin by deriving the number density of the discrete-arc comprised of point masses to accurately represent the continuous-arc. Based on the resonant characteristics of the 2:3 MMR, we then construct three overlapping discrete-arcs to model the Plutinos. The perturbations of these arcs are investigated in detail, considering various azimuthal and radial distributions associated with the resonant amplitudes $A$ and eccentricities $e$ of the Plutinos, respectively.}
   {The change in Sun-Neptune distance, i.e. $\Delta d_{SN}$, caused by Plutinos increases as the range of $A$ widens. At $e\lesssim0.1$, $\Delta d_{SN}$ can reach magnitudes on the order of 100 km. However, the effects of Plutinos' $A$ and $e$ can possibly balance each other. As given $e\gtrsim0.25$, we find that $\Delta d_{SN}$ approaches zero, indicating a negligible contribution from highly eccentric Plutinos to the planetary ephemerides. We finally provide a concise analytic expression, which contains the parameters $A$, $e$ and the total mass of Plutinos, to estimate $\Delta d_{SN}$ at any epoch from 2020 to 2120. Furthermore, since the difference in $\Delta d_{SN}$ between the arc and ring model can be as large as 170 km, we conclude that the ring model is unsuitable for representing the perturbations of Plutinos. The idea of the multiple-arc model designed for Plutinos can be readily generalized to other MMRs heavily populated by small bodies.}
   {}

	
	\maketitle

\section{Introduction}

Current planetary ephemerides can provide highly accurate predictions. However, when comparing the predicted positions of planets in the solar system with direct observations, certain discrepancies still remain. In a study by \citet{folkner2014planetary}, the authors addressed this issue by calibrating the orbital uncertainties of the planets based on observations. Using the Very Long Baseline Array observations of spacecraft at Mars, they showed that the orientation of the ephemerides is aligned with the International Celestial Reference Frame with an accuracy of 0".0002. This remarkable precision indicates that the errors in the orbits of terrestrial planets could be as small as a few hundred metres. Meanwhile, the positional accuracies of Jupiter and Saturn, derived from spacecraft tracking data, are in the order of tens of kilometers. Observations of the farthest planets, Uranus and Neptune, rely mainly on astronomical observations. Due to limitations posed by the Earth's atmosphere and the accuracy of star catalogs, the positional accuracies of Uranus and Neptune are currently at the level of a few thousand kilometers. Although such orbital errors may not be very small, further in situ observations could significantly improve them.

The development of planetary ephemerides is crucial for the space exploration. Traditionally, ephemeris computations account for Newtonian interactions between the Sun and planets, Sun oblateness J2, figure and tide effects, relativistic corrections, and lunar librations. However, with a vast number of asteroids observed in the solar system, modern planetary ephemerides have been updated to incorporate their gravitational perturbations \citep{folkner2014planetary, fienga2008inpop06, pitjeva2014development}. The contributions are mainly from three asteroid populations: main belt asteroids, Jupiter Trojans, and Kuiper Belt objects.

The main belt asteroids (MBAs) are distributed between the orbits of Mars and Jupiter, with a current estimated population of around 700,000. To simulate the global perturbation caused by the MBAs, they are commonly divided into two components: the `Bigs' and the `Smalls'. The Bigs refer to the MBAs that have the most significant influence on the Earth-Mars distance, and they are individually included in the numerical ephemerides calculations. Typically, about 300 or somewhat more Bigs are chosen for inclusion \citep{pitjeva2018mass, fienga2020inpop}. While the Smalls represent the remaining numerous MBAs and are modeled collectively as a homogeneous ring \citep{KRASINSKY200298, fienga2008inpop06, pitjeva2014development}. In the ephemeris INPOP06 \citep{fienga2008inpop06}, it is estimated that the perturbation induced by this asteroid ring on the Earth-Mars distance can reach up to 150 m during the time interval of 1969-2010. In order to reduce the time cost associated with the point-mass model, which involves a large quantity of individual MBAs, \citet{liu2022ring} proposed an alternative approach known as the multiple-ring model. They classified all MBAs, regardless of their sizes (i.e. including both the Bigs and Smalls), into over 100 families, and eventually assigned them to six rings with varying parameters. With regard to the global perturbation of the MBA, the six-ring model yields a Mars-Earth distance that deviates from the value obtained in the point-mass model by an error below 0.5 m over a 10-year period. The main advantage of this new approach is that it can improve computational efficiency compared to the previous `Bigs + ring' model, as there is no need for individually incorporating asteroids in the ephemeris calculations.

Jupiter Trojans (JTs) are asteroids that share the orbit of Jupiter, but leading and trailing Jupiter by about $60^{\circ}$ in longitude, i.e. around the L4 and L5 triangular Lagrangian points. These asteroids are said to be settled in the 1:1 mean motion resonance with Jupiter. As of February 2023, more than 12,000 JTs are registered in the Minor Planet Center (MPC)\footnote{http://www.minorplanetcenter.net/iau/lists/TNOs.html}. Since the L4 and L5 swarms are distributed in two separate regions, respectively, this unique configuration of JTs can not be simulated by a simple ring model like those used for the MBAs. Furthermore, there are much more JTs in the L4 swarm than the L5 swarm, and the number ratio is supposed to be between 1.3 and 2 \citep{jewi04, szabo2007properties, grav2011wise, grav2012wise, li2023JT}. The number asymmetry of JTs also contradicts the assumption of a homogeneous-ring model. In \citet{li2018constructing}, besides the 226 largest JTs with absolute magnitudes $H<11$, we modeled the remaining objects with $H>11$ using two arcs located around Jupiter's L4 and L5 points, respectively. Our findings demonstrate that the total effect of JTs can lead to a change of $\sim70$ m in the Earth-Mars distance during the 2014-2114 time interval.

The Kuiper belt objects (KBOs) are icy celestial bodies beyond the orbit of Neptune, and the majority of them are located in the region between 39.4 AU and 47.8 AU. Within this region, KBOs can be categorized into two distinct groups \citep{gladman2008nomenclature}: (1) Resonant KBOs: These objects occupy the mean motion resonances (MMRs) with Neptune, and they have small to large eccentricities ranging from $e=0.05$-0.35. A significant portion of resonant KBOs are found in Neptune's 3:2 MMR at $\sim39.4$ AU, sharing this resonance with Pluto. This particular population is usually called `Plutinos'. The second largest group of resonant KBOs reside in Neptune's 2:1 MMR at $\sim47.8$ AU. It is worth noting that these two MMRs correspond to the inner and outer boundaries for most of observed KBOs, as mentioned previously. (2) Classical KBOs: These objects are not in Neptune's MMRs and typically exhibit small to moderate eccentricities and inclinations. Estimations from \citet{bannister2016outer} indicate that the classical KBOs account for approximately half of the total mass of all KBOs.

To date, thousands of KBOs have been discovered, and their influence on the motion of the giant planets is warrants careful evaluation. Since the late 2000s, the perturbation model of KBOs in planetary ephemerides has been gradually refined. In the ephemeris EPM2008 developed by \citet{pitjeva_2009}, apart from the 21 biggest KBOs, the perturbations caused by the remaining smaller objects are modeled by a one-dimensional ring with a heliocentric radius of 43 AU. By adjusting the mass of the ring to fit the observation data obtained from spacecraft, she first estimates the total mass of KBOs. Later, in the ephemeris EPM2013, \citet{pitjeva2014development} improve the KBO model by considering the 31 most massive objects as individuals, while the smaller objects are still collectively represented by the same single ring. Considering that KBOs are primarily distributed between Neptune's 2:3 and 1:2 MMRs, \citet{pitjeva2018mass} introduce an 8 AU-wide annulus spanning from 39.4 AU to 47.8 AU to represent the global perturbation of numerous small KBOs. However, this two-dimensional ring model poses a severe drawback as the objects within the annulus rotate as a whole. To avoid this drawback, these authors propose a new model comprising three separate rings: two rings are positioned at 39.4 AU and 47.8AU, representing the 2:3 and 1:2 resonant KBOs, respectively; and the third ring is placed at 44 AU, symbolizing the `core' of the Kuiper belt predominantly inhabited by the classical KBOs. In a recent analysis by \citet{di2020analysis}, the same three-ring model as in \citet{pitjeva2018mass} is also employed to estimate the mass of the Kuiper belt, using the high-precision measurements of Saturn obtained from the Cassini mission.

When dealing with a large number of uniformly distributed asteroids, a single- or multiple-ring model may be appropriate for representing their global perturbation. However, considering the resonant KBOs, their motions are restricted to specific regions within the resonance's phase space, rather than encompassing it entirely \citep{li2022machine}. Therefore, there may be a certain level of inaccuracy when using a ring model to represent the global perturbation of these resonators. Previous studies by \citet{pitjeva2018mass} and \citet{di2020analysis} acknowledged that the KBOs residing in Neptune's 2:3 and 1:2 MMRs should be treated as individual populations. But this is far from sufficient to capture the distinctive features of the spatial distribution of resonant KBOs, as their most notable resonant motions driven by Neptune's MMRs have not been taken into consideration.

\begin{table*}
\centering
\begin{minipage}{13cm}
\caption{\label{table:1} Eleven most massive KBOs currently known.}
\begin{tabular}{l l l l}        
\hline                 

Number & Name & Mass $(10^{-4}M_{\oplus}$) & Reference                  \\

\hline 
136199 & Eris & $27.96\pm0.33$ & \citet{brown2007mass}\\
134340 & Pluto + Charon & $24.47\pm0.11$ & \citet{brozovic2015orbits}\\
136108 & Haumea & $6.708\pm0.067$ & \citet{ragozzine2009orbits}\\
136472 & Makemake & $4.35\pm0.84$ & \citet{pitjeva2018mass}\\
225088 & 2007 OR10 & $2.93\pm0.117$ & \citet{kiss2019mass}\\
50000 & Quaoar & $1.67\pm0.17$ & \citet{fraser2013mass}\\
90482 & Orcus & $1.0589\pm0.0084$ & \citet{brown2010size}\\
208996 & 2003 AZ84 & $0.69\pm0.33$ & \citet{pitjeva2018mass}\\
120347 & Salacia & $0.733\pm0.027$ & \citet{stansberry2012physical}\\
174567 & Varda & $0.446\pm0.011$ & \citet{grundy2015mutual}\\
55637 & 2022 UX25 & $0.2093\pm0.0050$ & \citet{brown2013density}\\
\hline
\end{tabular}
\end{minipage}
\end{table*}

In this paper, our focus is on the Plutinos, which are the largest resonant population observed in the Kuiper belt. The resonant angle $\sigma$ associated with Neptune's 2:3 MMR is defined by 
\begin {equation}
\sigma=3\lambda-2\lambda_N-\varpi,
\label{eq:resonantargument}
\end {equation}
where $\lambda$ and $\varpi$ are the mean longitude and the longitude of perihelion of the Plutino, respectively, and $\lambda_N$ is the mean longitude of Neptune. For a stable Plutino, its resonant angle $\sigma$ librates around $180^{\circ}$ with a resonant amplitude of $A<180^{\circ}$. Consequently, $\sigma$ can not traverse from 0 to $360^{\circ}$. This indicates that the mean longitudes of Plutinos are not evenly spread across the range of 0-$360^{\circ}$. Our primary objective in this study is to develop the planetary ephemerides that account for this asymmetry in the azimuthal distribution of Plutinos.

In our previous study on the perturbation of JTs \citep{li2018constructing}, a similar issue regarding the azimuthal distribution was discussed. In that study, the JTs are allowed to be on circular orbits with eccentricities $e=0$, because the resonant term of the 1:1 Jovian MMR in the expansion of the disturbing function does not contain $e$. However, Neptune's 2:3 resonance is the eccentricity-type, and its strength is proportional to $e$. Consequently, Plutinos must have $e$ values greater than 0. When involving the contribution of $e$, their orbital configuration becomes more complex. Therefore, we need to additionally consider the resulting asymmetry in the radial distribution of Plutinos.

The rest of this paper is organised as follows. In Sect. \ref{sec:dym}, we describe the design of the dynamical model of the solar system, and discuss the option of using the Sun-Neptune distance as a measurement for the perturbation induced by the KBOs. In Sect. \ref{sec:perturbation}, we construct both a ring model and an arc model to simulate the global perturbation of the Plutinos, and subsequently compare the results obtained from these two models. The new arc model is comprised of three overlapping arcs. We investigate in detail the plausible azimuthal and radial distributions of the arcs, which are determined by Plutinos' resonant amplitudes $A$ and eccentricities $e$, respectively. And then we supply a concise analytic expression to describe the contributions of $A$, $e$, and the total mass of Plutinos to the change in the Sun-Neptune distance. Finally, the conclusions and discussion are given in Sect. \ref{sec:conclu}. 

\section{Dynamical model of the solar system}
\label{sec:dym}

The unperturbed model of the solar system comprises the Sun and eight planets from Mercury to Neptune. The planets’ masses, initial heliocentric positions, and velocities are taken from DE405 \citep{standish1998jpl}. First, their orbits are adjusted from the mean equatorial system to the J2000.0 ecliptic system at epoch 2021 July 5. From this epoch forward, we proceed to construct the perturbed model of the solar system by incorporating the gravitational perturbations of massive KBOs. In the subsequent analysis, our objective is to evaluate the influence of the KBOs on the modern planetary ephemerides. To achieve this, we compare the motion of Neptune, the closest planet to the KBOs, in both the unperturbed and perturbed models.

To quantify the gravitational effect of KBOs on the orbit of Neptune, we examine the variation in the distance between the Sun and Neptune, defined as
\begin{equation}
 \Delta d_{SN}=d_{SN1}-d_{SN0},
 \label{dsn}
\end{equation}
where $d_{SN1}$ and $d_{SN0}$ indicate the Sun-Neptune distances calculated with and without considering the perturbations from the KBOs, respectively. In our numerical simulations of the solar system's evolution, we utilize the 19th-order Cowell prediction-correction algorithm (PECE) with a time-step of 0.5 day, which is chosen based on the orbital period of the innermost celestial body (i.e. Mercury) in our models \citep{tian1993adams, li2018constructing}. This N-body code employed in our calculations takes into account the gravitational interactions among the Sun, the planets, and the KBOs, but we neglect the gravitational forces between the KBOs themselves.

\begin{figure}
  \resizebox{\hsize}{!}{\includegraphics{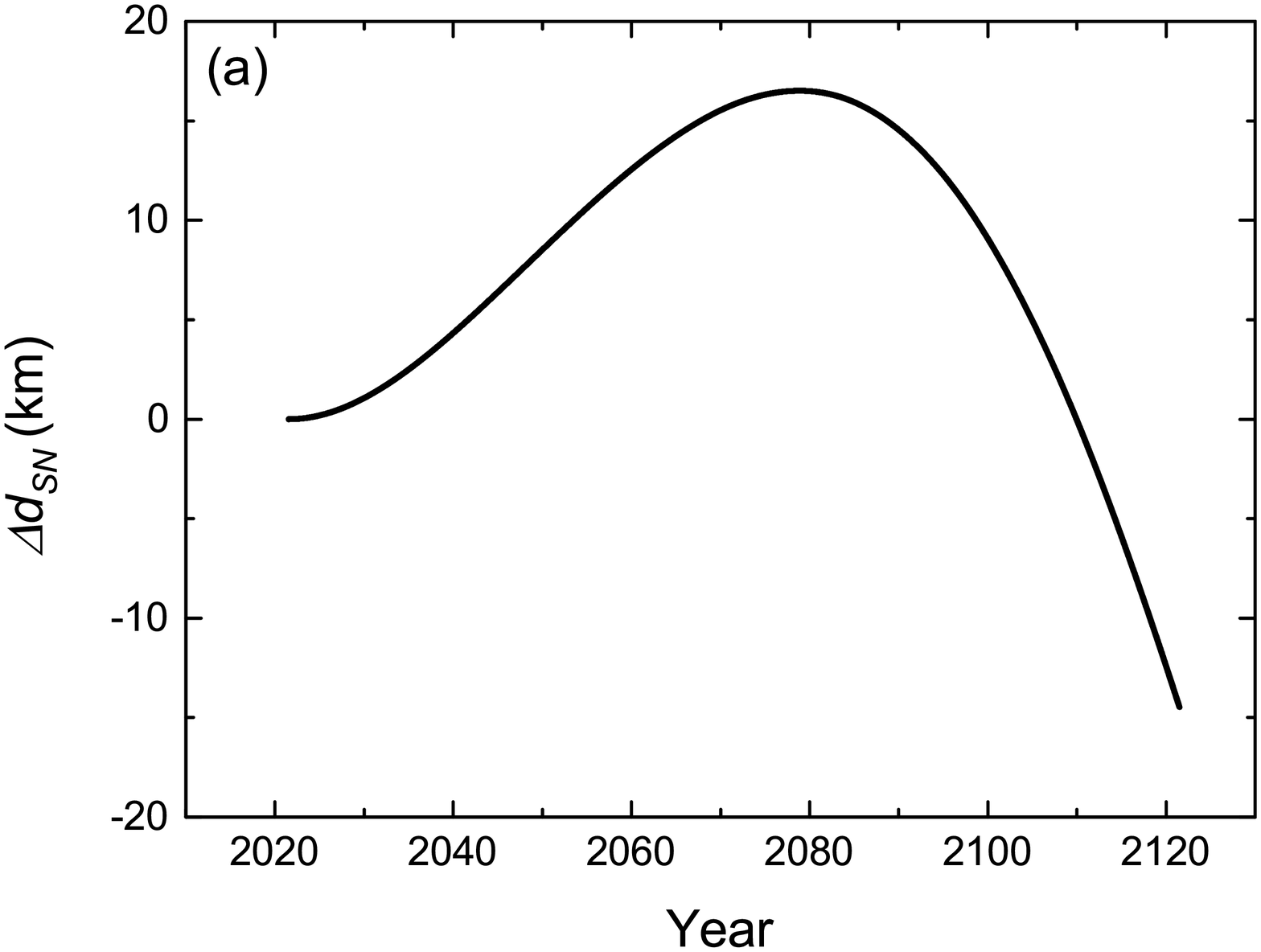}}
  \resizebox{\hsize}{!}{\includegraphics{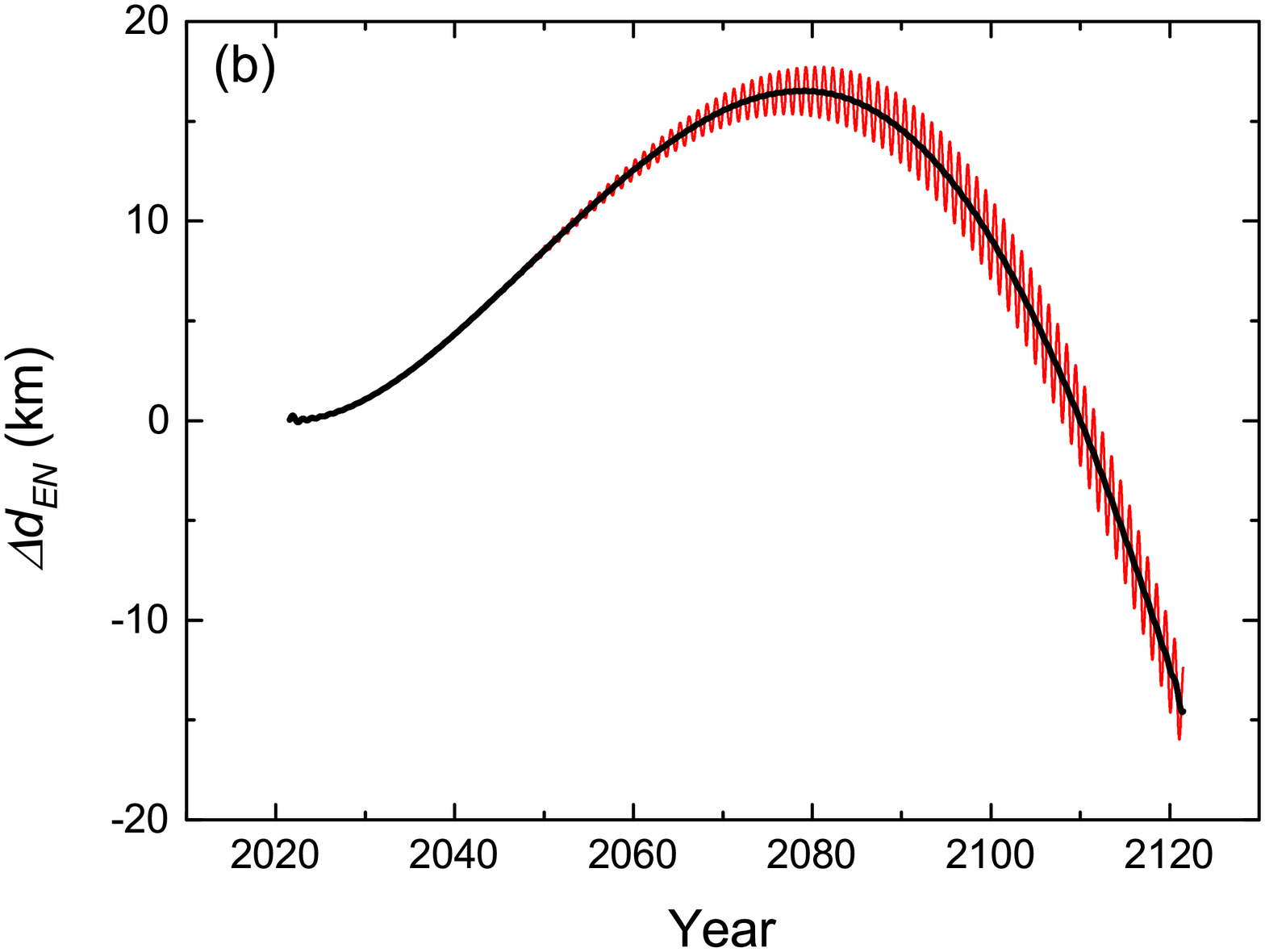}}
  \caption{Perturbation on the position of Neptune induced by the 11 most massive KBOs in the time interval between years 2020 and 2120: (a) the change in the Sun-Neptune distance, denoted as $\Delta d_{SN}$; (b) the change in the Earth-Neptune distance, denoted as $\Delta d_{EN}$. In this panel, the red curve represents the original $\Delta d_{EN}$, while the black curve depicts the secular behaviour of $\Delta d_{EN}$.}
  \label{fig:1}
\end{figure}

First of all, we begin by estimating the magnitude of the perturbation caused by the KBO population on the Sun-Neptune distance. For the sake of simplicity, we here take into account the influence of the 11 most massive KBOs. As listed in Table \ref{table:1}, their masses have been accurately determined through extensive research. While the orbital elements of these objects are gathered from the MPC, at the specific epoch of 2021 July 5 as we mentioned above. This way, we can calculate the perturbations from these 11 prominent KBOs, as measured by the change in the Sun-Neptune distance, i.e. $\Delta d_{SN}$. Figure \ref{fig:1}a shows that between the years 2021 and 2121, the resultant $\Delta d_{SN}$ could reach an approximate value of 16.5 km. This considerable value indicates that the perturbation induced by the KBOs on the Sun-Neptune distance is indeed significant. Consequently, it is imperative to develop modern planetary ephemerides to account for their gravitational effects.

It is important to note that the 11 selected objects represent only a fraction of the entire population of the KBOs. These 11 objects possess a total mass of about $7\times10^{-3} M_{\oplus}$, where $M_{\oplus}$ denotes the Earth mass. However, the total mass of the complete KBO population could be an order of magnitude larger, reaching up to $0.2M_{\oplus}$, as estimated theoretically by \citet{pitjeva2018mass}. Therefore, it is reasonable to suppose that the collective influence of all the KBOs exerted on Neptune's position could be even stronger. Additionally, the spatial distribution of the KBOs will significantly impact the magnitude of $\Delta d_{SN}$, as we will discuss later when considering the Plutinos.

In this paper, we intend to use the Sun-Neptune distance to measure the perturbation of the KBOs. However, due to the fact that the current observations are typically conducted from Earth or its vicinity, it is more common to determine the position of a planet relative to Earth \citep{standish1998jpl,folkner2014planetary,fienga2020inpop}. For example, the Earth-Mars distance is usually utilized to measure the perturbations induced by the MBAs \citep{fienga2008inpop06} and JTs \citep{li2018constructing}. To validate our choice of using the change in the Sun-Neptune distance, considering or excluding the 11 massive KBOs listed in Table \ref{table:1}, we also calculate the change in the Earth-Neptune distance, defined by
\begin{equation}
 \Delta d_{EN}=d_{EN1}-d_{EN0},
\end{equation}
where the variables have the similar meanings to Eq. (\ref{dsn}), but the subscript `E' indicates Earth. In Fig. \ref{fig:1}b, the red curve illustrates the temporal variation of $\Delta d_{EN}$ resulting from the perturbation of these 11 KBOs. Over a period of 100 yr, the change in the Earth-Neptune distance can reach a value as large as $\Delta d_{EN}\sim17.7$ km, which is comparable to the largest $\Delta d_{SN}$ displayed in Fig. \ref{fig:1}a.

Moreover, upon examining the red $\Delta d_{EN}$-curve shown in Fig. \ref{fig:1}b, one can observe the short-period oscillations occurring on the timescale of 1 yr, corresponding to the Earth's revolution around the Sun. To extract the low-frequency variation in the Earth-Neptune distance, we applied a fast Fourier transform (FFT) low-pass filter to $\Delta d_{EN}$. This process yields the smooth black curve, as plotted in Figure \ref{fig:1}b. By comparing the black curves in Figs. \ref{fig:1}a and \ref{fig:1}b, we find a remarkable agreement between $\Delta d_{SN}$ and the smoothed $\Delta d_{EN}$, as the difference between these two quantities is at a level of less than 0.4 km. Based on the results obtained here and above, we propose that the Sun-Neptune distance can serve as a representative measurement for evaluating the gravitational perturbations caused by the KBOs. 

One advantage of adopting the Sun-Neptune distance as a metric is its simplicity in describing the secular variation of Neptune's position over time, as we can effectively avoid the short-period variations observed in the Earth-Neptune distance due to the Earth's annual movement. More importantly, the other notable advantage is that employing the Sun-Neptune distance can considerably reduce the computational cost. Without the need to provide the Earth's position in locating Neptune, we can remove the four terrestrial planets from the dynamical model of the solar system. Consequently, Jupiter becomes the innermost planet instead of Mercury. According to the Jupiter-to-Mercury period ratio of $\sim20$, we can increase the integration time-step from 0.5 day to 10 days \citep{li2007origin}, thus the numerical integration process would be substantially faster.

In the simplified model of the solar system, the gravitational effects of the terrestrial planets are implemented by adding their combined masses to the Sun. However, before proceeding with further investigations, we have to validate that this approximation does not substantially influence the change in the Sun-Neptune distance caused by the perturbations of the KBOs. To assess this, we again calculate the value of $\Delta d_{SN}$ resulting from the perturbations of the 11 most massive KBOs within this simplified model. The results reveal that the profile of the time evolution of $\Delta d_{SN}$ remains nearly identical to that obtained in the original model, which includes all eight planets. Indeed, the largest $\Delta d_{SN}$ reaches a value of 16.5 km, which is only different by a relative error of 0.019\% compared to the original model. Therefore, the simplified solar system model allows us to reduce the calculation expense to merely 1/20 of the original amount without compromising our main results.

In a brief summary, our dynamical model for subsequent calculations comprises the Sun, four giant planets, and optionally includes the KBOs. Additionally, we will focus on a specific group of KBOs, i.e. the Plutinos in the 2:3 MMR with Neptune. To evaluate the gravitational perturbation exerted by these resonant KBOs, we will use the change in the Sun-Neptune distance, i.e. $\Delta d_{SN}$. 

\section{Perturbation of the Plutinos}
\label{sec:perturbation}

When considering a large number of KBOs uniformly distributed in azimuth, their global perturbation should be represented by a continuous ring. However, for the sake of computational simplicity, it is common to approximate this continuous ring with a discrete one, which consists of moving point masses. In \citet{pitjeva2018mass} and \citet{di2020analysis}, they adopted 40 point masses to represent the discrete ring. Although the number seems very small, it is still reasonable because the masses and positions of these point masses can be adjusted to fit observational data. But our current study aims to theoretically investigate the difference between the ring and arc models, so we have to minimize the impact of such discretization. 

From a geometrical perspective, as the number $n$ of point masses increases, the discrete ring approximation becomes increasingly accurate in representing the continuous ring. Similarly, for the arc model under investigation, we can consider its number density in relation to the value of $n$. Determining an appropriate number of point masses within a ring or arc can help us achieve a balance between computational efficiency and maintaining the robustness of our simulations. Furthermore, the use of a discrete arc model provides another advantage. By assigning eccentricities to the orbits of the point masses in the discrete model, we can readily account for the influence of eccentric KBOs.

In the following investigation of both the ring and arc models, we only focus on the zero-inclination case. This approach aligns with previous works that modeled KBOs using the one-dimensional ring \citep{pitjeva_2009} or the two-dimensional annulus \citep{pitjeva2018mass}. In particular, it is reasonable to consider Plutinos with orbital inclinations of $i=0$, as $i$ has limited significance for Neptune's 2:3 resonance, which is the eccentricity-type.

\subsection{The ring model}
\label{subsec:ring}

For a continuous ring, its gravitational potential exerting at the location $(x',y', z')$ can be described as
\begin{equation}
  \begin{aligned}
	&U(x',y',z',t)=-\int_r\int_0^{2\pi}\frac{G\rho_ar}{dist}drd\phi \\
	&dist=\sqrt{(rcos\phi-x')^2+(rsin\phi-y')^2+z'^2}\label{eq:ring}
  \end{aligned}
\end{equation}
where $G$ is the gravitational constant, $\rho_a$ is the linear density of the ring, and $(r,\phi)$ are the polar coordinates of the ring relative to the Sun. Notably, the two parameters $\rho_a$ and $r$ are assumed to be fixed, resulting in Eq. (\ref{eq:ring}) with only one variable, i.e. $\phi$. Consequently, the integration can be performed using the Romberg algorithm.

In order to determine the linear density $\rho_a$ of the ring, it is necessary to obtain the total mass of the Plutinos. The total mass of the entire population of KBOs was estimated to be in a large range of $M_{kb}=0.01-0.2M_\oplus$ \citep{pitjeva2018mass}. Later, by analysing high-precision measurements of Saturn from the Cassini mission, \citet{di2020analysis} suggested a more specific total mass of $M_{kb}=0.061M_\oplus$ for the Kuiper belt population, with Plutinos accounting for approximately $1/6$ of this mass. Based on these findings, we will assume a total mass of $M_{plu}=0.01M_\oplus$ for the Plutinos. It should be noted that the mass $M_{plu}$ still remains quite uncertain. Nevertheless, to first-order accuracy, the perturbation caused by the Plutinos is linearly proportional to $M_{plu}$ \citep{li2018constructing}. Therefore, our subsequent analyses could be nearly independent of the specific value chosen for $M_{plu}$, as we will discuss in Section \ref{result}.

Regarding the radial distance $r$ of the ring, we should choose it to be the nominal 2:3 resonance location of 39.4 AU. However, this choice may induce a potential issue associated with the discrete representation of the ring using point masses. If the point masses were distributed around 39.4 AU, the uniformity of the discrete ring would be affected by the 2:3 resonance. In order to avoid this issue when comparing the continuous and discrete ring models, we select a slightly larger radial distance of $r=43.4$ AU, which is 1.44 times Neptune's semi-major axis of $a_N=30.1$ AU. At the heliocentric distance of 43.4 AU, objects with eccentricities smaller than 0.1 are not in the 8th or lower order resonances of Neptune \citep{li2023study}.

\begin{figure}
  \resizebox{\hsize}{!}{\includegraphics{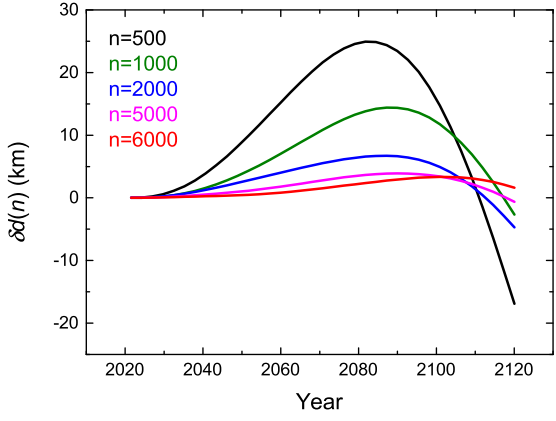}}
  \caption{The difference in $\Delta d_{SN}$ between the continuous and discrete ring models, that is, $\delta d(n)$ defined in Eq. (\ref{Dring}). The variable $n$ indicates the number of point masses used to construct a discrete ring.}
  \label{fig:2a}
\end{figure}

\begin{figure}
    \resizebox{\hsize}{!}{\includegraphics{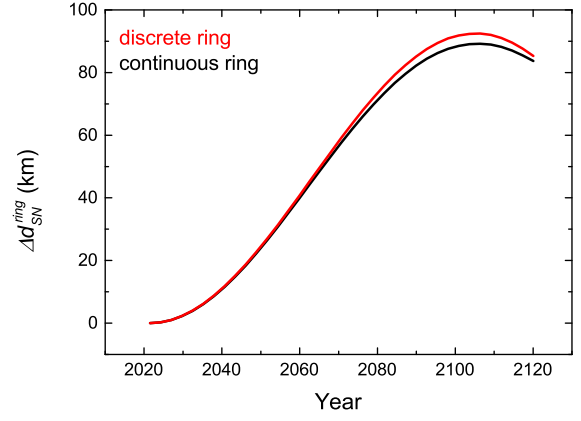}}
    \caption{Perturbations on the Sun-Neptune distance induced by the continuous ring (black) and the discrete ring comprised of 6000 point masses (red). This figure corresponds to the specific case of $n=6000$ shown in Fig. \ref{fig:2a}.}
    \label{fig:2b}
\end{figure}

We then proceed to calculate the changes in the Sun-Neptune distance induced by both the continuous and discrete rings, denoted as $\Delta d_{SN}^{ring}$ and $\Delta d_{SN}^{ring}(n)$, respectively. The difference between the perturbations of these two rings can be expressed as:
\begin{equation}
  \delta d(n)=\Delta d_{SN}^{ring}(n)-\Delta d_{SN}^{ring},
\label{Dring}
\end{equation}
where $n$ refers to the number of point masses used to construct the discrete ring. As we discussed earlier, increasing the value of $n$ helps reduce the effects of discretization, quantified by $\delta d(n)$ in Eq. (\ref{Dring}). We compute the values of $\delta d$ by varying $n$ from 500 to 6000, and the results are shown in Fig. \ref{fig:2a}. It is evident that the difference $\delta d(n)$ exhibits temporal variation due to the evolution of the system. Therefore, we parameterize this difference by its maximum value, denoted as $\max \delta d(n)$, within the considered epoch from 2020 to 2120. For $n=500$, the value of $\max \delta d(n)$ can be as large as about 30 km. However, as $n$ increases to 5000, $\max \delta d(n)$ decreases by an order of magnitude, reaching only 3.8 km. In addition, with a further increase in $n$ to 6000, we observe that $\max \delta d(n)$ changes at a much smaller level of only $\sim0.1$ km. Thus we suppose that $\delta d(n)$ has nearly converged to a small value of $<4$ km, and it is unlikely to change significantly during the 100 yr time period.

To better visualize the small discrepancy between the perturbations induced by the continuous and discrete rings, we focus on the case where $n=6000$. Figure \ref{fig:2b} depicts the temporal variations of $\Delta d_{SN}^{ring}$ and $\Delta d_{SN}^{ring}(n)$, represented by the black and red curves, respectively. We find that the relative error between these two curves remains below 4\% throughout the 100-yr period. This implies that $n=6000$ can be considered an appropriate number of point masses for constructing the discrete ring, to achieve a reliable approximation of the continuous ring.

By setting $n=6000$ for a discrete ring, its number density $\Sigma_0$ in azimuth can be calculated via 
\begin{equation}
    \Sigma_0=\frac{6000}{2\pi}\frac{r}{r_0},
\label{number_ring}
\end{equation}
where $2\pi$ is the radian measure of a complete ring, $r$ is the heliocentric radius of the ring, and $r_0=43.4$ AU is the reference radius. In the subsequent investigation of the arc model, it is essential to ensure that the number density of the discrete arc satisfies the condition given in Eq. (\ref{number_ring}). Let $l$ be the length of a discrete arc in azimuth, the minimum number of the point masses required to construct this arc can be calculated as:
\begin{equation}
    n_{arc}=\Sigma_0 \times l.
\label{number_arc}
\end{equation}
Then for the arc-like longitude distribution of Plutinos driven by Neptune's 2:3 resonance, we will first determine the associated length $l$ in the upcoming section.

\subsection{The three-arc model}
\label{subsec:arc}

As the primary concern of this paper, it is important to note that Plutinos, a distinct group of KBOs trapped in Neptune's 2:3 resonance, cannot be mimicked as a homogeneous ring due to their unique spatial distribution. We now are about to determine the azimuthal distribution of Plutinos in physical space. For the 2:3 resonance, its critical resonant angle $\sigma$ is defined by Eq. (\ref{eq:resonantargument}). This resonance has a stable equilibrium point at $\sigma_0=180^{\circ}$, around which $\sigma$ undergoes libration with a resonant amplitude $A<180^{\circ}$. According to the expression for $\sigma$, the azimuthal distribution of Plutinos can be represented by the differences in the mean longitude between the Plutinos and Neptune, as: 
\begin{equation}
\Delta\lambda=\lambda-\lambda_N=\frac{1}{3}(\sigma+\varpi-\lambda_N).
\label{longitude}
\end{equation}

\begin{figure}
  \resizebox{\hsize}{!}{\includegraphics{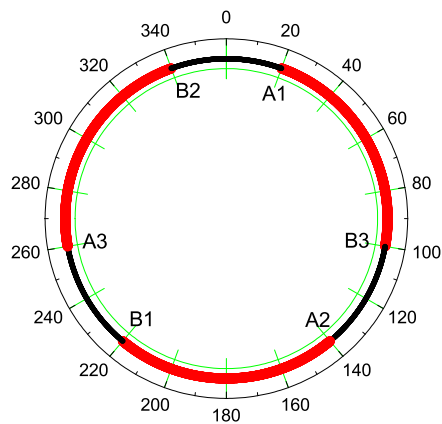}}
  \caption{Schematic diagram of Plutinos' azimuthal distribution (the unit is degree), for the case of the resonance amplitudes $A\le 120^{\circ}$. This diagram shows three overlapping Plutino-arcs: A$1$-B$1$, A$2$-B$2$, and A$3$-B$3$. Each arc represents the combination of two adjacent red curves with a black curve between them.}
  \label{fig:plu}
\end{figure}

To establish the initial conditions for the Plutinos, we select the resonant angles $\sigma$ uniformly from $180^{\circ}-A_{max}$ to $180^{\circ}+A_{max}$, where $A_{max}$ represents their maximum resonant amplitude; and, the longitudes of perihelia $\varpi$ are randomly chosen between 0 and $360^{\circ}$. Then, given Neptune's mean longitude $\lambda_N$ at the beginning of the calculation, we can determine the initial mean longitudes $\lambda$ of the Plutinos from Eq. (\ref{longitude}). Additionally, it must be noted that when we unfold the phase space of the 3:2 resonance, three resonant islands emerge, corresponding to $\sigma\in$ [0, $360^{\circ}$], [$-360^{\circ}$, 0] and [$360^{\circ}$, $720^{\circ}$], respectively \citep{li2023study}. As a result, apart from the commonly considered range of [0, $360^{\circ}$], $\sigma$ is allowed to vary within the other two ranges by associating the angle $\sigma=\sigma\pm360^{\circ}$. While these displacements in $\sigma$ preserve the resonant behaviours of the Plutinos, the measurement of their longitude positions relative to Neptune, i.e. $\Delta\lambda$, would significantly change. For example, when we set the resonant amplitudes to be $A \le A_{max}=120^{\circ}$ and give an initial $\lambda_N=0$, the initial values of $\lambda$ are distributed within three intervals of $[20^{\circ}, 220^{\circ}]$, $[140^{\circ}, 340^{\circ}]$, and $[260^{\circ}, 360^{\circ}]\bigcup[0, 100^{\circ}]$. Figure \ref{fig:plu} sketches the resulting azimuthal distribution of Plutinos, where the three arcs (A$1$-B$1$, A$2$-B$2$, and A$3$-B$3$) corresponds to the three respective $\lambda$ intervals just given. These arcs overlap each other at the regions indicated by the red thicker curves. This configuration could be analogous to Neptune's bright arcs, which actually are the concentrations of particles embedded within the Adams ring \citep{pater2018}. It is worth noting that within the red regions, the number density of Plutinos is twice that of the black regions. So obviously, the perturbation of Plutinos should be described using the three-arc model instead of the ring model.

For Plutinos with zero-inclination orbits, besides $\lambda$ and $\varpi$, two additional orbital elements need to be set: the semimajor axis $a$ and the eccentricity $e$. We put all Plutinos initially at $a=39.4$ AU, which corresponds to the nominal location of Neptune's 2:3 resonance; and, their initial $e$ values can be varying. Bearing in mind that the initial $\lambda$ is derived from the resonant angle $\sigma$ via Eq. (\ref{longitude}), and we have $\sigma\in [180^{\circ}-A_{max}, 180^{\circ}+A_{max}]$, these Plutinos form three distinct arcs as illustrated in Fig. \ref{fig:plu}. The azimuthal lengths of these arcs are determined by $A_{max}$, while their curvatures are influenced by the value of $e$. Thus, the maximum resonant amplitude $A_{max}$ and the eccentricity $e$ of Plutinos serve as the two adjustable parameters of our three-arc model. According to the orbits of observed Plutinos \citep{li2014study}, we choose representative values for these two parameters within the ranges of $A_{max}=10^{\circ}$-$120^{\circ}$ and $e=0.05$-0.35. The upper limit of $A_{max}=120^{\circ}$ is chosen because Plutinos with $A\le120^{\circ}$ typically exhibit stable resonant orbits over the age of the solar system. As for the lower limit of $e=0.05$, since the strength of the 2:3 resonance is proportional to $e$, the objects with $e<0.05$ are generally too weak to sustain their resonant behaviours. As a matter of fact, $A_{max}$ and $e$ can respectively introduce asymmetries in the azimuthal and radial distributions of Plutinos, resulting in different magnitudes of changes in the Sun-Neptune distance. More interesting, the effects of these two orbital parameters on the spatial distribution asymmetry of Plutinos can possibly balance each other, as we will show below.


To ensure an adequate number of Plutinos in each discrete arc for them to effectively replace the continuous arc, we consider the case of $A_{max}=120^{\circ}$. In this scenario, the azimuthal length of each arc reaches its maximum of $200^{\circ}$, as depicted by the individual arcs A$i$-B$i$ ($i=1,2,3$) in Fig. \ref{fig:plu}. Let $l=200^{\circ}=10\pi/9$ and $r=39.4$ AU in Eqs. (\ref{number_ring}) and (\ref{number_arc}), we can obtain that a minimum number of $n_{arc}\approx3000$ Plutinos are needed for representing a single arc. Accordingly, in subsequent 100-yr calculations, we employ 9000 equal-mass Plutinos (i.e. point masses) for our three-arc model, regardless of $A_{max}$. The total mass of these Plutinos is still adopted to be $M_{plu}=0.01M_\oplus$, as what we did for the ring model. We want to mention that, the distribution of Plutinos within the arcs may vary during the 100 yr evolution. Nevertheless, this timescale is extremely shorter than the 2:3 resonance's period of 20,000 yr, leading to very small variation of $\sigma$ ($<3.6^{\circ}$). Therefore, the azimuthal distribution of Plutinos can be reasonably considered as nearly consistent throughout the entire computational duration. This analysis supports our choice of 9000 Plutinos to effectively approximate the continuous arcs.


\subsection{Perturbation difference between the ring and arc models}
\label{result}


To compare the global perturbation of Plutinos mimicked by the ring and arc models, we calculate the corresponding changes $\Delta d_{SN}$ in the Sun-Neptune distance. In the ring model, $\Delta d_{SN}$ is derived from the continuous ring described by Eq. (\ref{eq:ring}), where the radius $r$ is fixed at 39.4 AU. On the other hand, in the arc model, $\Delta d_{SN}$ is obtained from three discrete arcs that we constructed in Sect. 3.2. These arcs are also located at 39.4 AU, but have varying orbital parameters $A_{max}$ and $e$ to consider the possible distribution of observed Plutinos.

\begin{figure}
  \centering
  \begin{minipage}[c]{0.5\textwidth}
  \centering
  \vspace{-0.3cm}
  \includegraphics[width=8.5cm]{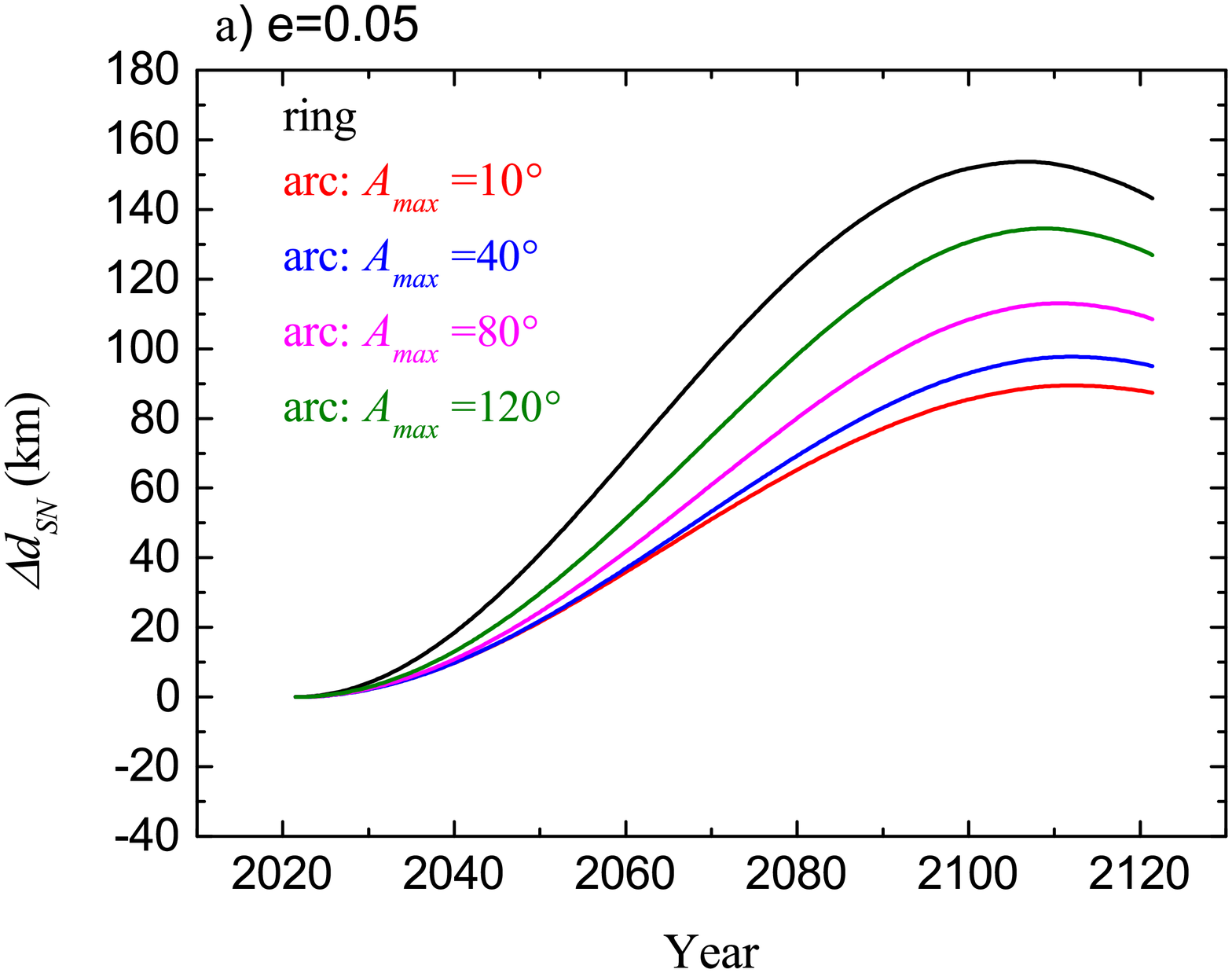}
  \end{minipage}
  \begin{minipage}[c]{0.5\textwidth}
  \centering
  \vspace{-0.6cm}
  \includegraphics[width=8.5cm]{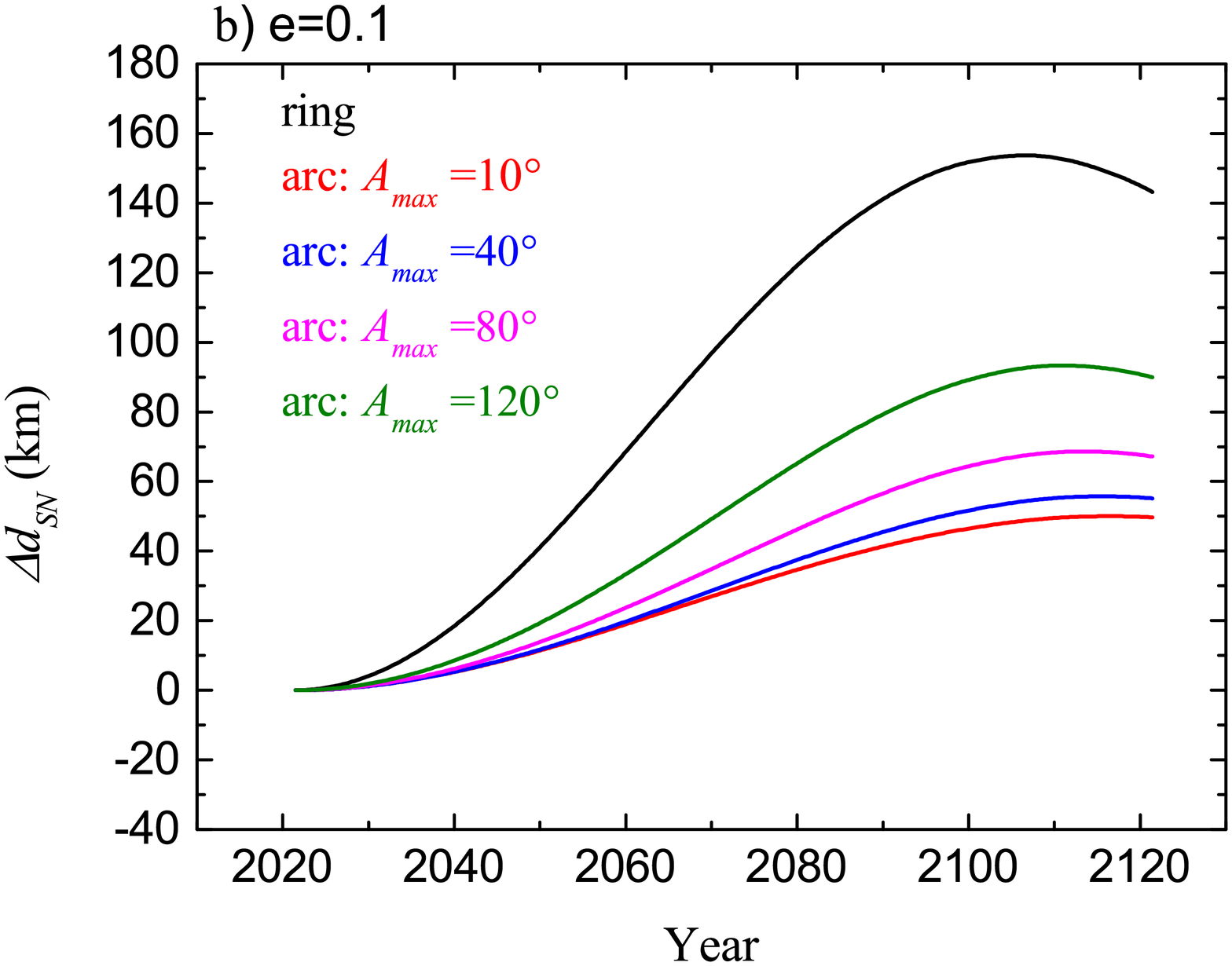}
  \end{minipage}
  \begin{minipage}[c]{0.5\textwidth}
  \centering
  \vspace{-0.6cm}
  \includegraphics[width=8.5cm]{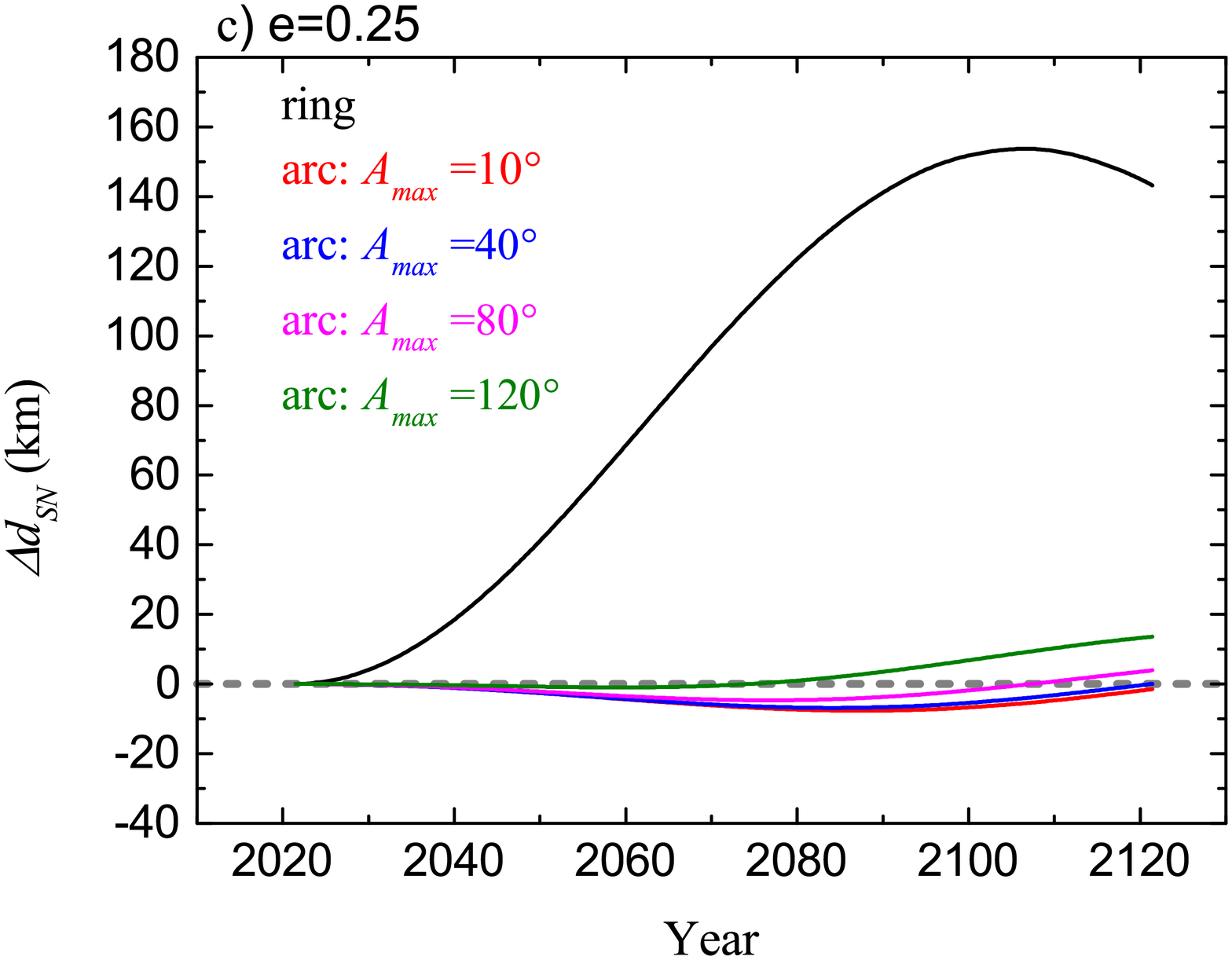}
  \end{minipage}
  \begin{minipage}[c]{0.5\textwidth}
  \centering
  \vspace{-0.6cm}
  \includegraphics[width=8.5cm]{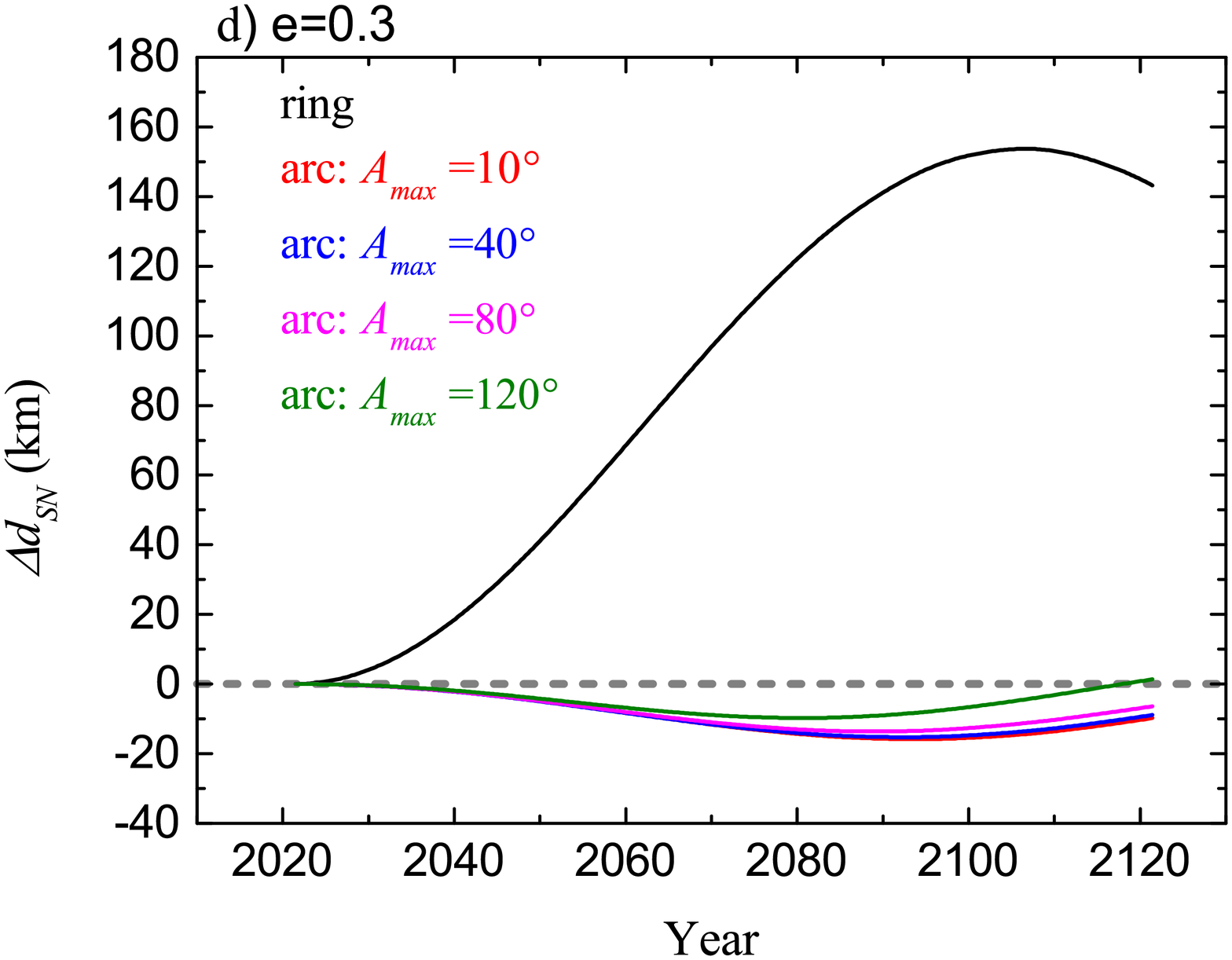}
  \end{minipage}
    \caption{Perturbation on the Sun-Neptune distance induced by the three arcs, within which Plutinos possess resonant amplitudes of $A \le A_{max}$ and varying eccentricities, as: (a) $e=0.05$, (b) $e=0.1$, (c) $e=0.25$, and (d) $e=0.3$. The black curve in each panel represents the same reference case of a continuous ring fixed at 39.4 AU. In the two lower panels, a horizontal dashed line is plotted at $\Delta d_{SN}=0$, which corresponds to the unperturbed solar system model.}
  \label{fig:3}
\end{figure}

The temporal evolutions of $\Delta d_{SN}$ are plotted in Fig. \ref{fig:3}, covering a time span of 100 yr from 2020 to 2120. In each panel (a)-(d), we present the same $\Delta d_{SN}$ variation for the ring model (depicted by the black curve), which will be considered as the reference case. It is observed that the change in the Sun-Neptune distance, induced by the Plutino-ring perturbation, experiences a substantial increase right from the beginning in 2020, reaching a maximum value of $\Delta d_{SN}\sim154$ km around the year 2100. Subsequently, $\Delta d_{SN}$ slightly decreases to about 143 km by the end of our calculation at the epoch 2120.

Next, we examine the three-arc model starting by the case of $e=0.05$, as shown in Fig. \ref{fig:3}a. In order to gain a clearer view of the variation trend in $\Delta d_{SN}$ resulting from changes in the azimuthal distribution of Plutinos, we selectively plot color-coded curves for representative scenarios of $A \le A_{max}=10^{\circ}$ (red), $40^{\circ}$ (blue), $80^{\circ}$ (magenta), and $120^{\circ}$ (green). For each given $A_{max}$, the Plutinos have their resonant angles $\sigma$ confined to the range of $[180^{\circ}-A_{max}, 180^{\circ}-A_{max}]$. One can see that for Plutinos with resonant amplitudes $A \le A_{max}=10^{\circ}$, around the year 2100, $\Delta d_{SN}$ reaches its peak of 89 km, which is about half of the maximum $\Delta d_{SN}$ obtained in the ring model (indicated by the black curve). But as $A_{max}$ increases from $10^{\circ}$ to $120^{\circ}$, we find that the corresponding curve (from red to green) is approaching the black one associated with the ring model, suggesting a diminishing asymmetry in the perturbation of Plutinos. This phenomenon can be attributed to the fact that the $\sigma$-range widens at larger $A_{max}$, and becomes closer to the complete interval of $\sigma=[0, 360^{\circ}]$ observed in the ring model. Nevertheless, even when $A_{max}=120^{\circ}$, the discrepancy in $\Delta d_{SN}$ between the ring (black curve) and arc (green) models can still exceed 10 per cent.

Figure \ref{fig:3}b illustrates the outcomes for another representative case with a small $e=0.1$. We can see that the difference between the $\Delta d_{SN}$ values derived from the ring and arc models gradually decreases as $A_{max}$ increases. This trend in $\Delta d_{SN}$ with increasing $A_{max}$ is still apparent and similar to the case of $e=0.05$ (see Fig. \ref{fig:3}a), indicating a less pronounced asymmetry in the azimuthal distribution of Plutinos.

However, when the eccentricities of Plutinos are large, the asymmetry in their radial distribution becomes important and can have a substantial impact on the Sun-Neptune distance. Figures \ref{fig:3}c and d show the results for the cases of $e=0.25$ and 0.3, respectively. Remarkably, a compelling pattern emerges as the colorful curves representing different $A_{max}$ values cluster around $\Delta d_{SN}=0$. These high-eccentricity cases serve as a notable highlight of our arc model in two key respects: (1) the azimuthal and radial distribution asymmetries of Plutinos determined by $A_{max}$ and $e$, respectively, can possibly cancel each other out. In such situations, the perturbation caused by Plutinos becomes exceptionally weak. The resulting effect on the Sun-Neptune distance differs by only 10-20 km over the entire 100-yr evolution compared to the unperturbed solar system model, where $\Delta d_{SN}=0$. (2) Regardless of $A_{max}$, the colourful curves deviate prominently from the black curve, exhibiting differences as large as about 170 km in $\Delta d_{SN}$. This significant deviation clearly demonstrates that the ring model is completely unsuitable for accurately representing the perturbations induced by the Plutinos.

\begin{figure}
  \resizebox{\hsize}{!}{\includegraphics{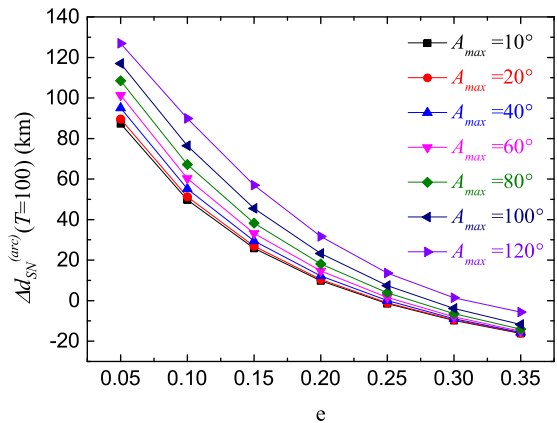}}
  \caption{Dependence of $\Delta_{SN}$ on the eccentricities $e$ of Plutinos for seven different $A_{max}$ values: $10^{\circ}$ (black), $20^{\circ}$ (red), $40^{\circ}$ (blue), $60^{\circ}$ (magenta), $80^{\circ}$ (green), $100^{\circ}$ (navy), and $120^{\circ}$ (purple). The values of $\Delta_{SN}$ are generated using the three-arc model at the end of our 100-yr calculations (i.e. at the epoch 2120).}
  \label{fig:5}
\end{figure}

Finally, we provide a concise description for the dependence of the perturbation of Plutinos on their parameters: the maximum resonant amplitude $A_{max}$, the eccentricity $e$, and the total mass $M_{plu}$. Figure \ref{fig:5} summarizes the changes in the Sun-Neptune distance induced by the three-arc model at the end of our 100-yr calculation, denoted as $\Delta d^{(arc)}_{SN}(T=100)$, for various values of $A_{max}$ and $e$. Through linear fitting of the data points corresponding to different $(e, A_{max})$ combinations in this figure, we establish the following measurement: 
\begin{equation}
    \Delta d^{(arc)}_{SN}(T=100)=\alpha e+\beta ~~~\mbox{(km)},
    \label{eq:deltadarc}
\end{equation}
where the coefficients $\alpha$ and $\beta$ are given by
\begin{equation}
\begin{split}
    \alpha&=-310.6 (A_{max} / 1^{\circ})-1.1, \\
    \beta&=78.7 (A_{max} / 1^{\circ})+0.4.
\end{split}
\end{equation}
In addition, as depicted in Fig. \ref{fig:3}, the temporal variation of the $\Delta d_{SN}$ (i.e. $\Delta d^{(arc)}_{SN}$) can be approximated as linear. Therefore, we refine Eq. (\ref{eq:deltadarc}) to account for any specific epoch $T$ between the years 2020 ($T=0$) and 2120 ($T=100$) as follows:
\begin{equation}
    \Delta d^{(arc)}_{SN}(T)=(\alpha e+\beta)\frac{T}{100} ~~~\mbox{(km)}.
    \label{eq:deltadarc}
\end{equation} 

Furthermore, the total mass $M_{plu}$ of Plutinos may be a crucial parameter in determining $\Delta d^{(arc)}_{SN}$. Although the precise value of $M_{plu}$ currently remains uncertain, the first-order approximation made in \citet{li2018constructing} suggests that the perturbations induced by point masses are nearly proportional to their masses. By incorporating the mass factor into Eq. (\ref{eq:deltadarc}), we can derive a comprehensive expression for the perturbation of Plutinos on the Sun-Neptune distance:
\begin{equation}
\Delta d^{(arc)}_{SN}=\frac{M_{plu}}{0.01M_\oplus}(\alpha e+\beta)\frac{T}{100} ~~~\mbox{(km)}.
\label{eq:AeMt}
\end{equation}
At this point, our arc model is decoupled from the parameter $M_{plu}$, which can be significantly improved as future surveys increase the real Plutino sample in number.

\section{Conclusions and discussion}
\label{sec:conclu}

To achieve highly accurate planetary ephemerides, it is crucial to refine the solar system model by incorporating the influence of numerous asteroids. In the previous work \citep{li2018constructing}, we began to consider the perturbations induced by the Jupiter Trojan asteroids. As these asteroids reside in the 1:1 MMR with Jupiter, they do not exhibit a uniform distribution in azimuth and can not be simply mimicked by a homogeneous ring. Therefore, we developed a model using two separate arcs centered at Jupiter' L4 and L5 points, respectively. Continuing our investigation into the perturbations of the resonant asteroids, this second paper is devoted to Plutinos, which are the KBOs trapped in the 2:3 MMR with Neptune. We are aware that the 2:3 MMR entails three stable equilibrium points, indicating that Plutinos should also exhibit uneven azimuthal distribution.

In our dynamical model, the unperturbed solar system consists of the Sun and the four giant planets from Jupiter to Neptune, while the four terrestrial planets are implemented by adding their masses to the Sun. Additionally, to assess the impact of perturbations caused by the KBOs, we choose to measure the change in the Sun-Neptune distance, denoted as $\Delta d_{SN}$. In order to account for the perturbation effects, we first consider the gravitational influence of the 11 largest KBOs as individual bodies within the solar system. This approach allows us to validate the adoption of the simplified solar system model and the use of $\Delta d_{SN}$ as a measurement. The advantages of these choices have also been presented, particularly in terms of computational efficiency.

In order to accurately simulate the total perturbation of Plutinos, we need to employ a sufficiently large number of point masses that are distributed uniformly in azimuth but limited to a specific range rather than spanning the entire 0-$360^{\circ}$. To accomplish this task, we evaluate the value of $\Delta d_{SN}$ by considering the perturbations generated by either a continuous ring model represented by its gravitational potential (Eq. (\ref{eq:ring})), or a discrete ring model comprised of $n$ individual moving point masses. As the number $n$ increases, the discrepancy in $\Delta d_{SN}$ between these two models diminishes. When setting $n\ge6000$, this discrepancy remains below 4\% over a 100-yr calculation. Hence, we deduce that a minimum number density of $6000/2\pi$ per radian in the azimuth is required for a discrete ring (or arc) to effectively replace a continuous one, ensuring an accurate representation of the gravitational perturbation of the latter.

Next, we analyse the possible azimuthal distribution of Plutinos resulting from the libration of the resonant angle $\sigma$ of Neptune's 2:3 MMR. Observations indicate resonant amplitudes $A$ that do not exceed $120^{\circ}$. Consequently, $\sigma$ is allowed to vary from $60^{\circ}$ to $300^{\circ}$, excluding the full 0 to $360^{\circ}$ range. As a result, the mean longitudes $\lambda$ of Plutinos relative to that of Neptune fall with the following intervals:  $[20^{\circ}, 220^{\circ}]$, $[140^{\circ}, 340^{\circ}]$, and $[260^{\circ}, 360^{\circ}]\bigcup[0, 100^{\circ}]$. To capture this azimuthal distribution, we construct three overlapping arcs to represent the perturbations induced by Plutinos. Each arc's length is equal and determined by the maximum resonant amplitude $A_{max}$, which could be $120^{\circ}$ as an upper limit. Subsequently, the number of point masses within each arc is calculated using the obtained minimum number density, resulting in about 3000 point masses per arc. Hence, our three-arc model comprises a total of 9000 Plutinos, each having equal mass. We consider $A_{\text{max}}$ as an adjustable parameter, and as it decreases below $120^{\circ}$, the lengths of the three arcs become smaller, leading to severer asymmetry in the azimuthal distribution of Plutinos.

We proceed by evaluating the perturbations induced by the three arcs, assuming zero inclinations. For Plutinos within these arcs, the mean longitudes $\lambda$ are derived from the resonant angles $\sigma$ corresponding to different resonant amplitudes $A$, where $A\le A_{max}$. We set their semi-major axes to be $a=39.4$ AU, which is the nominal location of Neptune's 2:3 MMR, and the longitudes of perihelia $\varpi$ are randomly selected between 0 and $360^{\circ}$. As for their eccentricities $e$, we start by considering the lower limit of $e=0.05$ based on current observations. We then calculate the cumulative effect of these Plutinos on the change of the Sun-Neptune distance, i.e. $\Delta d_{SN}$, during the 2020-2120 time interval. We find that $\Delta d_{SN}$ induced by the three arcs monotonously increases with $A_{max}$, reaching a peak of 134 km at the largest $A_{max}=120^{\circ}$. While in the reference case of a continuous ring fixed at 39.4 AU, the resulting $\Delta d_{SN}$ can be up to $\sim154$ km. Therefore, for the case of $e=0.05$, the relative difference in generating $\Delta d_{SN}$ between the arc and ring models exceeds $10\%$ over the 100-yr period under consideration.

As a matter of fact, the azimuthal distribution of Plutino-arcs is controlled by $A_{max}$, while their radial distribution is determined by $e$. It is intriguing to discover that these two parameters have a balancing effect on the perturbation of Plutinos. As $e$ increases, the difference in $\Delta d_{SN}$ between the arc and ring model becomes more prominent. Notably, when $e \gtrsim 0.25$, we emphasize two key findings: (1) The perturbation caused by Plutinos on the Sun-Neptune distance becomes significantly small, as $\Delta d_{SN}$ reaches only 10-20 km within the considered epoch from 2020 to 2120. (2) Regardless of the resonant amplitudes of Plutinos, the $\Delta d_{SN}$ value obtained in the arc model can deviate by as much as 170 km from that in the ring model. Thus, the ring model is clearly inadequate for accurately representing the perturbation induced by Plutinos.

To summarize the above results, we finally provide a concise analytic expression in Eq. (\ref{eq:AeMt}) to estimate the change in the Sun-Neptune distance caused by Plutinos. This expression takes into account the resonant amplitudes, eccentricities, and total mass of Plutinos, and any specific epoch between the years 2020 and 2120.


Furthermore, it should be noted that in this theoretical study, although the arcs are confined to specific ranges that are narrow than 0-$360^{\circ}$, they are evenly distributed in azimuth. However, due to the limited number of currently observed Plutinos, only around 400, it is not feasible to conduct a comprehensive statistical analysis. As a result, the Plutinos may be sparse in certain parts of the arcs while being dense in others. This potential inhomogeneity in the distribution of Plutinos would further diminish the applicability of the ring model. Nonetheless, our arc model offers a significant advantage in that we can readily address this issue by considering a series of sub-arcs, each with different $A_{max}$ values, for the three main arcs.


In recent times, various projects for surveying the KBOs have been proposed, for example, the ecliptic Deep Drilling Field by the Large Synoptic Survey Telescope (LSST) \citep{jone16, trilling2018deep}. So, a larger number of small and faint KBOs are to be discovered, enabling us to impose even tighter constraints on the spatial distribution and the mass of Plutinos. Consequently, a finer model of KBOs can be achieved and it can help us to enhance the accuracy of planetary ephemerides even further.


\begin{acknowledgements}
    
This work was supported by the National Natural Science Foundation of China (Nos. 11973027, 11933001), and National Key R\&D Program of China (2019YFA0706601). 
      
\end{acknowledgements}

%
%

\end{document}